\documentstyle[11pt,newpasp,twoside,epsf]{article} 
\begin{document}  
 
\title{SuperMassive Black Holes from the SUNNS survey}
\author{Marc Sarzi} 
\affil{Dipartimento di Astronomia, Universit\`a di Padova, Vicolo
dell'Osservatorio 5, I-35122 Padova, Italy}
\author{Hans-Walter Rix}
\affil{Max-Planck-Institut f{\"u}r Astronomie, K{\"o}nigstuhl 17,
Heidelberg, D-69117, Germany}
\author{Joseph C. Shields}  
\affil{Physics \& Astronomy Department, Ohio University, Athens, OH
45701}
\author{Gregory Rudnick, Daniel H. McIntosh} 
\affil{Steward Observatory, University of Arizona, Tucson, AZ 85721}
\author{Luis C. Ho} 
\affil{The Observatories of the Carnegie Institution of Washington,
813 Santa Barbara St., Pasadena, CA 91101-1292}
\author{Alexei V. Filippenko} 
\affil{Astronomy Department, University of California, Berkeley, CA
94720-3411}
\author{Wallace L. W. Sargent} 
\affil{Palomar Observatory, Caltech 105-24, Pasadena, CA 91125}

\rule{0.5mm}{0cm} 
 
\noindent 
Over the last few years evidence has mounted that supermassive black
holes (SMBH) are nearly ubiquitous in galactic centers (see, e.g., Ho
1999 for a review).
The possibilty that a close connection between the formation and
evolution of SMBHs and of their host galaxies exists, has recently
been bolstered by reports of a correlation between the black-hole mass 
$M_{\bullet}$ and the velocity dispersion of the stellar bulge $\sigma$ 
(Ferrarese \& Merritt 2000; Gebhardt et al. 2000; Merritt \& Ferrarese 2000).
Despite this recent progress, the demography of SMBHs is far from
complete. Most of the SMBHs have been detected in bulge-dominated
systems, and it is not clear if the $M_{\bullet}$-$\sigma$ relation
holds also in the case of disk-dominated galaxies, or when a bar is
present.

\noindent
In order to address this and other issues we initiated a Survey of
Nearby Nuclei with STIS (SUNNS) on {\sl HST}, with kinematic
information obtained from measurements of nebular emission lines for
24 galaxies.
From this starting sample, four galaxies displayed symmetric gas
velocity curves that could be consistently modeled as a rotating disk
in the joint potential of the stellar bulge and a putative central
black hole.
An extensive description of the analysis for these galaxies can be
found in Sarzi et al. (2000).
In Figure 1 we show how the three sources for which we obtained
well-determined values of $M_{\bullet}$, namely NGC~2787{\it (SB0)},
NGC~4459{\it (S0)}, NGC~4596{\it (SB0)}, fall in the
$M_{\bullet}$-$\sigma$ plane.

\begin{figure}
\plotone{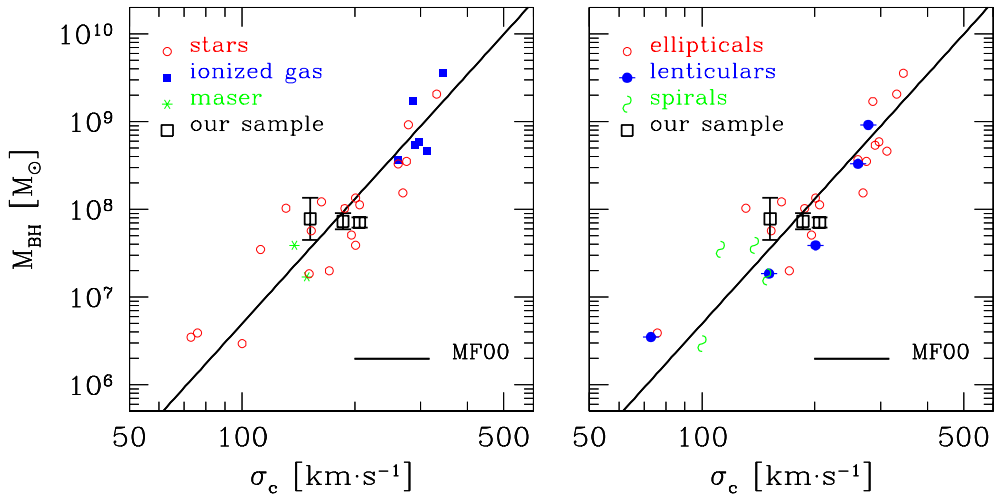} 
\caption{{\it Left:} Our 3 $M_{\bullet}$ determinations (open
squares) along with some of the $M_{\bullet}$ measurements from which
the Merritt \& Ferrarese (2000) relation (solid line) was derived
($M_{\bullet}$ from reverberation mapping are not shown). 
Different symbols correspond to the different tracer of the potential
used to obtain $M_{\bullet}$. {\it Right:} Same as {\it left} except
that different symbols now indicate the host galaxy Hubble-type.}
\end{figure} 

\noindent
Reliable $M_{\bullet}$ determinations based on gas kinematics have
been limited so far to ionized gas measurements in ellipticals and
water-maser measurements in late-type spirals.
Our $M_{\bullet}$ measurements fill the corresponding gap in the
$M_{\bullet}$-$\sigma$ plane, and our SUNNS survey illustrates both
the difficulty in finding regular gas motions in disk galaxies and the
need for an accurate preselection of the targets.
Finally our $M_{\bullet}$ determinations are not only consistent with
the $M_{\bullet}$-$\sigma$ relation, but also reinforce it.  
Indeed the number of disk galaxies with accurate $M_{\bullet}$
measurements (excluding reverberation mapping measurements) has now
been increased from 9 to 12, the number of S0's from 5 to 8, and the
number of barred galaxies has been doubled.

\end{document}